\documentclass[preprint,3p,times,twocolumn,longtitle]{elsarticle}

\usepackage{graphicx}

\usepackage{amssymb,amstext}

\usepackage{lineno}

\biboptions{sort&compress}

\usepackage{xspace}
\usepackage{pstricks}
\usepackage{color}	

\usepackage[%
  breaklinks=true,%
  colorlinks=true,%
  linkcolor=blue,anchorcolor=blue,%
  citecolor=blue,filecolor=blue,%
  menucolor=blue,pagecolor=blue,%
  urlcolor=blue]{hyperref}
\journal{Physics Letters B}

\newcommand{\hermes}{{\sc Hermes}\xspace}

\newcommand{\plotje}[2]{ \includegraphics[width = #1]{#2}  }

\newcommand{\pT}{P_{T}}

\begin{document}

\begin{frontmatter}




\title{
  Transverse target single-spin asymmetry
 in inclusive  electroproduction of charged pions and kaons\\
}


\author[13,16]{A.~Airapetian}
\author[27]{N.~Akopov}
\author[6]{Z.~Akopov}
\author[7]{E.C.~Aschenauer\fnref{28}}
\author[26]{W.~Augustyniak}
\author[27]{R.~Avakian}
\author[27]{A.~Avetissian}
\author[6]{E.~Avetisyan}
\author[19]{S.~Belostotski}
\author[11]{N.~Bianchi}
% ...

\fntext[28]{Now at: Brookhaven National Laboratory, Upton, New York 11772-5000, USA}

% ...
\address[6]{DESY, 22603 Hamburg, Germany}
\address[7]{DESY, 15738 Zeuthen, Germany}
% ...
\address[11]{Istituto Nazionale di Fisica Nucleare, Laboratori Nazionali di Frascati, 00044 Frascati, Italy}
% ...
\address[13]{II. Physikalisches Institut, Justus-Liebig-Universit\"at Gie\ss en, 35392 Gie\ss en, Germany
}
% ...
\address[16]{Randall Laboratory of Physics, University of Michigan, Ann Arbor, Michigan 48109-1040, USA }
% ...
\address[19]{K.P. Konstantinov Petersburg Nuclear Physics Institute, Gatchina, 188300 Leningrad Region, Russia}
% ...
\address[26]{National Centre for Nuclear Research, 00-689 Warsaw, Poland}
\address[27]{Yerevan Physics Institute, 375036 Yerevan, Armenia}

\begin{abstract}


Single-spin asymmetries were investigated in inclusive electroproduction 
of charged pions and kaons from transversely polarized protons
at the {\hermes} experiment. The asymmetries were studied as a function of  
the azimuthal angle $\psi$ about the beam direction between the 
target-spin direction and the hadron production plane, the transverse 
hadron momentum $\pT$ relative to the direction of the incident beam, 
and the Feynman variable $x_F$. 
The $\sin\psi$ amplitudes are positive for $\pi^{+}$ and $K^{+}$, slightly 
negative for $\pi^{-}$ and consistent with zero for $K^{-}$, with 
particular $P_{T}$ but weak $x_{F}$ dependences.
Especially large asymmetries are observed for two small subsamples of 
events, where also the scattered electron was recorded by the 
spectrometer.

\end{abstract}


\end{frontmatter}



Transverse single-spin asymmetries (SSAs) 
observed in the azimuthal distributions of hadrons produced in high-energy
electromagnetic and hadronic reactions (where either the projectile or the target nucleon is 
polarized transversely to the beam direction) are a window for our understanding of 
the nucleon structure and the process of hadronization in the framework 
of quantum-chromodynamics (QCD). 
They originate from correlations of the transverse spin of the nucleon and/or the transverse 
spins of the quarks  with transverse quark 
momentum and could in models be related to spin-orbit effects 
and to the elusive orbital motion of partons within the nucleon.
Left-right cross-section asymmetries $A_N$ for the inclusive production of various hadrons in 
hadron-nucleon collisions have been 
measured over the past three decades by numerous experiments 
~\cite{Dick75,Klem76,Drago78,Antille80,Apokin90,Adams91pi0,Adams91pi+,
Adams92pi0,Adams92eta,Bravar95,Adams96,Bravar96,Abramov97,Adams98,Krueger99, Allgower02,star04,phenix05,Bazilevsky06,Togawa07,brahms07,brahms08,star08, brahms09,
phenix10} for center-of-mass energies in the range 4.9 - 500 GeV.  
Large values of $A_N$ were observed for single hadrons in $p^{\uparrow}p\rightarrow hX$ 
reactions at large transverse hadron momenta, $\pT$, and large positive values of $x_F$, exceeding 
$\vert A_N \vert = 0.4$ for charged pions. The Feynman
variable $x_F$ is defined as the ratio of the longitudinal hadron momentum $P_L$ along the beam
direction to its maximum possible value.
Transverse single-spin asymmetries have also been investigated
in semi-inclusive deep-inelastic lepton scattering, $lN^{\uparrow}\rightarrow l'hX$, 
from transversely polarized hydrogen~\cite{hermes05colsiv,
hermes09siv,hermes10col,compass10p,compass12cp,compass12sp},
deuterium~\cite{compass05d,compass06d,compass08d}, and $^3$He~\cite{clashe3} targets.  
Here, substantial azimuthal SSAs up to about 
0.1 have been observed for hydrogen targets. 
A review of experimental results can be found in Refs.~\cite{dalesioMurgia} 
and~\cite{Barone10}, together with an extended discussion on  contemporary theoretical work.

The large size of these single-spin asymmetries indicates 
the importance of effects beyond the standard leading-twist framework based on collinear factorization. 
One approach~\cite{anselmino94} is based on the use of 
parton distribution and fragmentation functions that are unintegrated in transverse momenta. 
In this approach, the asymmetries 
are caused mainly by two mechanisms: the 
Sivers~\cite{sivers90} and Collins~\cite{collins92} effects. The former is related to 
the transverse-momentum-dependent naive-time-reversal odd Sivers distribution function 
of unpolarized quarks with non-zero transverse momenta in a transversely polarized nucleon. 
The latter is related to the chiral-odd transversity distribution of transversely 
polarized quarks in a transversely polarized nucleon, in conjunction with 
the transverse-momentum-dependent chiral-odd Collins fragmentation function.
The other approach~\cite{qiusterman,kanazawa00a,kanazawa00b,koike02,koike03, qiu06} 
links collinear parton dynamics to higher-twist  
multiparton correlations. Again, two mechanisms 
dominate where either a 
twist-three chiral-odd fragmentation function couples to the transversity distribution, 
or where a twist-three chiral-even distribution function enters with the ordinary 
leading-twist unpolarized fragmentation function.
These approaches have different kinematic domains of validity, but with a region in common.
In the past 
it was believed that they succeeded in reproducing the existing measurements of $A_{N}$ 
in hadron-hadron reactions to a very large extent, 
and have been shown to be related to and consistent with each other in the kinematic 
region where they both apply~\cite{koike07}. 
Recently, however, a sign error was identified~\cite{Kang11a} that invalidates the 
good agreement observed earlier. Presently the situation is unsettled~\cite{Metz12}.

There exist several theoretical expectations for aspects of the SSAs in hadron electroproduction. Their validity depends 
on the relative magnitude of the three relevant scales $\Lambda_\text{QCD}$, $\pT$ and $Q$, where 
$\Lambda_\text{QCD} \cong 0.3$ GeV is the QCD scale parameter and
-$Q^2$ is the squared four-momentum of the virtual photon that mediates the lepton-nucleon scattering process:

\noindent i) Theory makes no reliable prediction for the kinematic region where both $ \pT$ and $Q$ are 
small and of order $\Lambda_\text{QCD}$; 

\noindent ii) The twist-three approach leads to a characteristic power suppression by $1/\pT$ for large $\pT$, 
provided $\pT$ is the largest scale in the process. 
For $\pT < 1$ GeV such power suppressions typically become less efficient;

\noindent iii) The Sivers and Collins effects become 
significant when   $Q^2 > \pT^2$ and $Q^2 \gg \Lambda^2_\text{QCD}$ and give a contribution
that is not $\pT$-suppressed. For large $Q^2$, the dominant contribution to the asymmetry 
should therefore come from the Sivers and Collins mechanisms. The SSAs measured in semi-inclusive DIS were, 
in fact, the basis for an extraction of the Sivers and transversity distribution functions and the Collins fragmentation function 
(see e.g., Refs.~\cite{Anselmino11,Anselmino09});

\noindent iv) It was shown that, for the kinematic regime $ Q \gg \pT \gg \Lambda_\text{QCD}$,
the descriptions in terms of the Sivers distribution function and of twist-three quark-gluon 
correlation functions become equivalent~\cite{jiqiu06} and that there also exists a kinematic region 
in which  a twist-three fragmentation function and the leading-twist
Collins fragmentation function can be mapped onto one another~\cite{Yuan09}. 
For $\pT^2 \sim Q^2$ one cannot make any quantitative theoretical statement about their connection.

A substantial number of theoretical predictions (see, e.g., 
Refs.~\cite{koike07,Ji06,Eguchi06a,Eguchi06b,bacchetta08,Beppu10,Kang11b,
koike02,anselmino10}) have not yet 
been confronted with experimental data. 
More data are required in a wider kinematic range
that covers transverse momenta as high as possible but also approaches $\pT$ values 
as small as $\Lambda_\text{QCD}$ 
for both $A_{N}$ in hadron-hadron reactions and SSAs in electroproduction of hadrons, $lp^{\uparrow}\rightarrow hX$. 
This Letter reports on the first measurement of azimuthal SSAs in inclusive 
electroproduction of charged pions and kaons off transversely polarized protons. 
It addresses a portion of this unexplored kinematic space.

The data reported here were collected during the period 2002 - 2005 with the 
{\sc Hermes} spectrometer \cite{HERMESspec}
using the 27.6 GeV lepton beam (electrons or positrons) incident 
upon a transversely nuclear-polarized gaseous hydrogen target internal to the {\sc Hera} 
lepton storage ring at {\sc Desy}.
The integrated luminosity of the data sample was approximately 
146 pb$^{-1}$. The average magnitude of the proton-polarization component perpendicular 
to the beam direction, $S_{T}$, was $0.713 \pm 0.063.$
The direction of the target-spin vector was reversed 
between the \textquotedblleft upward\textquotedblright ~and 
\textquotedblleft downward\textquotedblright ~directions at 1-3 minute intervals to minimize 
systematic effects, while both the nuclear polarization and the atomic fraction inside 
the target cell were measured continuously \cite{target}. The beam was longitudinally 
polarized and its helicity reversed every few months. A
helicity-balanced data sample was used  to obtain an effectively unpolarized beam.

Selected events had to contain at least one charged-hadron track, 
identified as either a pion or a kaon, within the angular acceptance of the spectrometer 
($\pm 170$ mrad horizontally and $\pm (40-140)$ mrad vertically) independent of whether 
there was also a scattered lepton in the acceptance or not.  
Hadrons were distinguished from leptons by using a transition-radiation detector, 
a scintillator pre-shower counter, and an electromagnetic calorimeter.
This resulted in a tiny lepton contamination in the hadron sample of less than 
0.1\%. Hadrons within the momentum range  2 -- 15 GeV were further identified using 
a dual-radiator ring-imaging Cherenkov detector \cite{HERMESrich}. This identification 
is based on a direct ray tracing algorithm that deduces the most probable particle types
from the event-level hit pattern of Cherenkov photons on the photomultiplier matrix \cite{HERMEScosphi}.

The trigger of the experiment was formed, for each detector half, by a coincidence of 
signals from a scintillation counter 
in front of the spectrometer magnet and 
from a  scintillator hodoscope and the pre-shower counter behind the 
magnet, spaced by ~1 m, with the requirement of an energy deposit greater than 1.4 GeV 
in the electromagnetic calorimeter.
The trigger was almost $100\%$ efficient for leptons with energies above threshold. The energy threshold of the 
calorimeter was low enough to trigger also on events with only charged hadrons and no leptons in its 
geometrical acceptance. In this case, the trigger efficiency was substantially smaller and depended on 
the hadron momentum $P_h$, as well as on the impact position and angle of the hadron track on 
the calorimeter surface and the hadron multiplicity in the event. Averaged over the 
hadron multiplicity, the trigger efficiency was about $40-45\%$ for hadron momenta greater 
than approximately 7 GeV and decreased smoothly with decreasing $P_h$ to about $15\%$ at 
$P_h \approx 2$ GeV.
In order not to bias the inclusive-hadron sample towards events with a coincident lepton in the
 detector acceptance, trigger-efficiency corrections dependent on the event topology 
 (e.g., additional lepton or further hadrons in the event) were applied.
In total, about $60 \cdot 10^6$  ($50  \cdot 10^6$) tracks of positively (negatively) charged pions and 
$5.1 \cdot 10^6$  ($2.8  \cdot 10^6$) tracks of positively (negatively) charged kaons were collected.  
These correspond to about $172\cdot 10^6$  ($142  \cdot 10^6$) positively (negatively) charged pions and 
$14.5 \cdot 10^6$  ($7.3  \cdot 10^6$) positively (negatively) charged kaons after trigger-efficiency correction 
(cf.~Tab.~\ref{counts}), which are used in all of the subsequent results.

 \begin{figure}[t]
\centering
\plotje{0.37\textwidth}{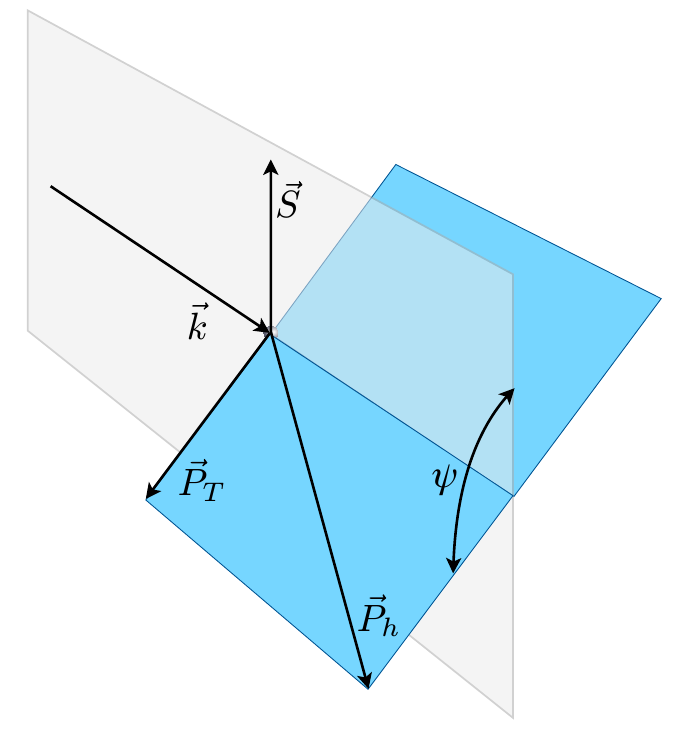} 
\caption{The definition of the azimuthal angle $\psi$. }
\label{fig:angle}
\end{figure}

As the scattered lepton was not required for  
the primary analysis, the following hadron variables were used:  $\pT$, the 
transverse momentum of the hadron with respect to the lepton beam direction; 
$x_F$, here calculated in the lepton-nucleon center-of-momentum frame; and $\psi$, the azimuthal angle about 
the beam direction between the \textquotedblleft upward\textquotedblright ~target 
spin direction and the hadron production 
plane, in accordance with the {\it Trento Conventions}~\cite{trento} 
(see Fig.~\ref{fig:angle}).

The cross section for inclusive electroproduction of hadrons 
using an unpolarized lepton beam and a transversely polarized target 
includes a polarization-averaged and a polarization-dependent part and is given for each 
hadron species as
\begin{equation}
\mathrm{d}\sigma
= \mathrm{d}\sigma_{UU}\left[1 + S_{T} A_{UT}^{\sin\psi}\sin\psi\right].
\label{eq:yields}
\end{equation}
Here, the first subscript U denotes unpolarized beam, the second subscript U (T) an unpolarized 
(transversely polarized) target. The dependences of the cross section and of the azimuthal amplitude 
$A_{UT}^{\sin\psi}$ on $\pT$ and $x_F$ have been omitted. 
The $\sin\psi$ azimuthal dependence follows directly from the 
term $\vec{S}\cdot (\vec{P}_h\times \vec{k})$ 
in the spin-dependent part of the cross section (see, e.g., Ref.~\cite{anselmino10}), 
with $\vec{S}$ being the target-spin vector, and $\vec{k}$ and $\vec{P}_h$ the three-momenta 
of the incident lepton and of the final-state hadron, respectively.

The $\sin\psi$ amplitude $A^{\sin\psi}_{UT}$ is related to the left-right asymmetry 
$A_{N}$ along the direction of the incident lepton beam and with respect to the nucleon-spin 
direction,\footnote{The sign convention of $A_{N}$ in hadron collisions commonly differs 
through defining ``left''  and ``right'' with respect to the momentum and transverse-spin 
directions of the incoming polarized hadron.} 
measured with a detector with full $2\pi$-coverage in $\psi$ and constant efficiency, by 

\noindent
\begin{equation}
A_N \equiv \frac{\int\nolimits_{\pi}^{2\pi}d\psi\,d\sigma - \int\nolimits_{0}^{\pi}d\psi\,d\sigma}{\int\nolimits_{\pi}^{2\pi}d\psi\,d\sigma + \int\nolimits_{0}^{\pi}d\psi\,d\sigma}= - \frac{2}{\pi} \ A^{\sin\psi}_{UT}.
\label{eq:anaut}
\end{equation}

Experimentally, the $A_{UT}^{\sin\psi}$ amplitudes 
were extracted by performing a maximum-likelihood fit to the 
cross section of Eq.~\ref{eq:yields}, i.e., the measured yield distribution
for the two target-spin states weighted with the inverse of the trigger efficiencies and luminosity,
binned in $\pT$ and $x_F$, but unbinned in $\psi$. 
The detection efficiency,
if independent of the target-spin state, 
cancels in the fit as long as the polarization-weighted 
luminosity vanishes, i.e., $\int \!\! S_{T}(t) L(t) dt = 0$, as 
is the case for the present data.

\begin{figure}[t]
\centering
\plotje{0.47\textwidth}{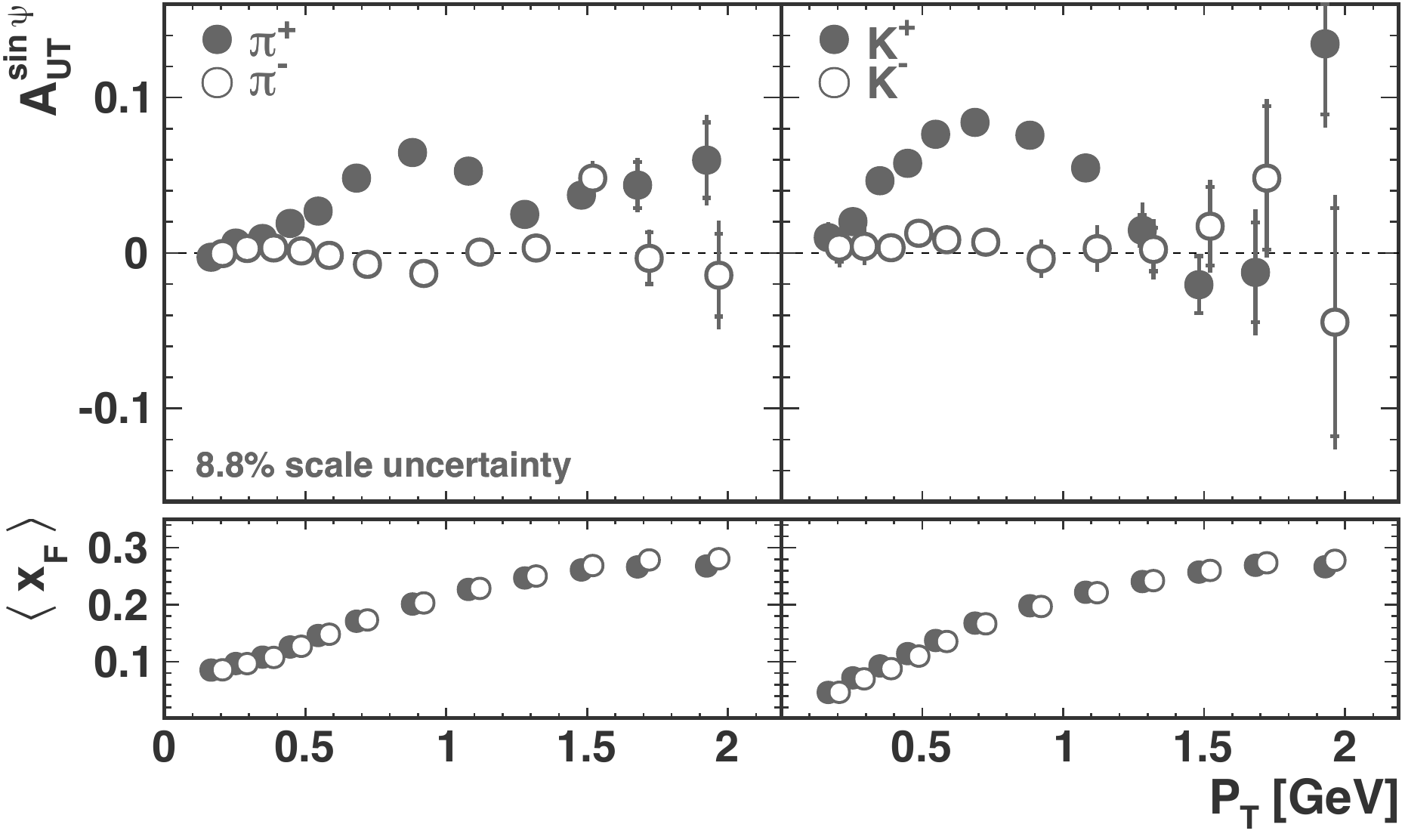} \\[2mm]
\plotje{0.47\textwidth}{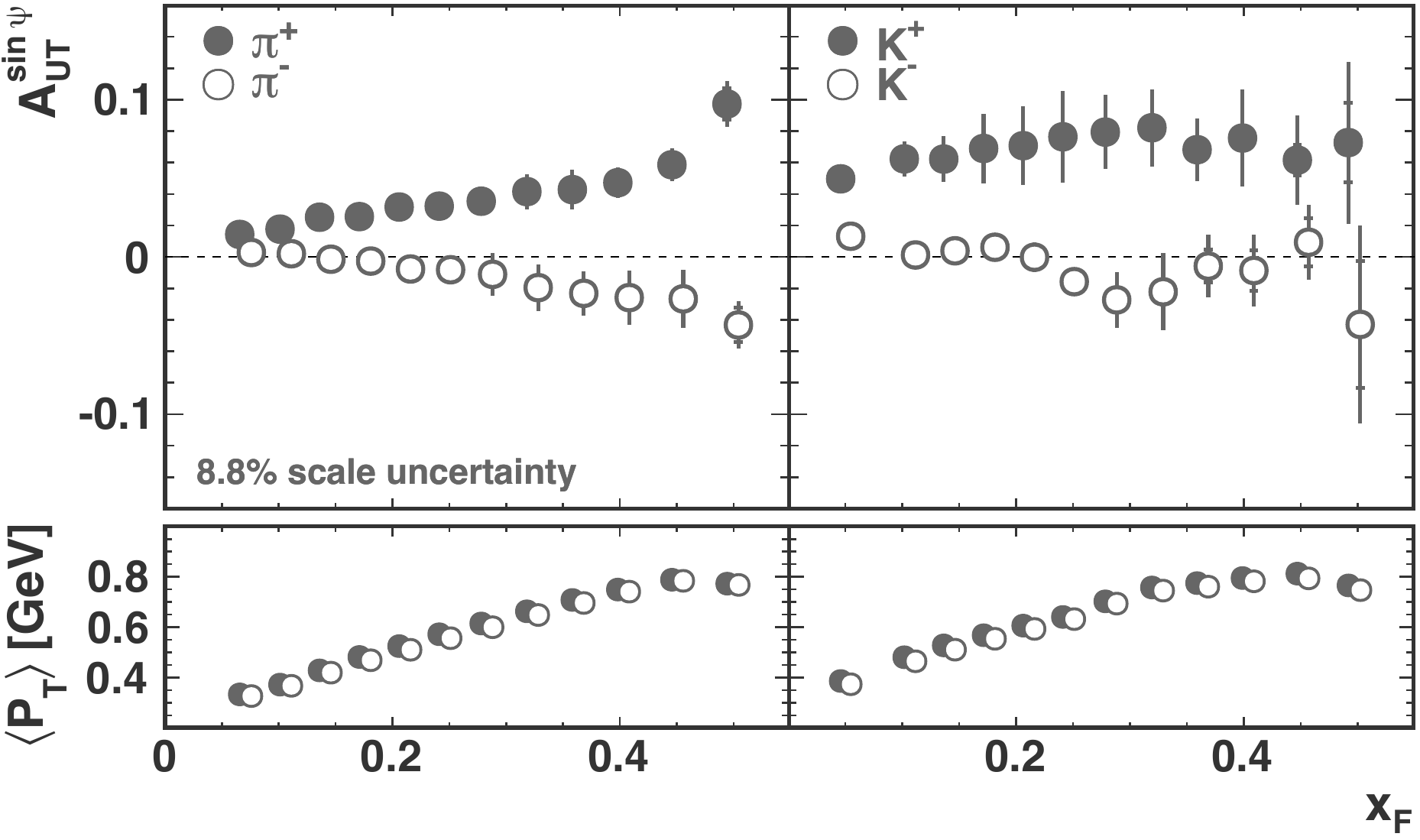}
\caption{$A_{UT}^{\sin\psi}$ amplitudes for charged pions and kaons as a function of 
$\pT$~(top) and $x_F$~(bottom). Positive (negative) particles are denoted by closed (open) symbols. 
When visible, the inner error bars show the statistical uncertainties, while the total ones represent the 
quadratic sum of statistical and systematic uncertainties. 
Not shown is an additional $8.8\%$  scale uncertainty due to the precision of the measurement of the target polarization.  
The bottom subpanels show the $\pT$ dependence ($x_F$ dependence) of the average $x_F$
($\pT$).
Data points for negative particles are slightly shifted horizontally for legibility.
}
\label{fig:asym}
\end{figure}

The extracted $A_{UT}^{\sin\psi}$ amplitudes for charged pions and kaons are presented 
as a function of $\pT$ in the top panels of
Fig.~\ref{fig:asym}.  The amplitudes are positive for the positive 
hadrons, being slightly larger for kaons compared to pions. They rise smoothly with $\pT$
up to a maximum value of approximately 
0.06 (0.08) for pions (kaons) at $\pT \simeq 0.8$~GeV and then decrease again with 
increasing $\pT$. Note that at $\pT = 0$~GeV the amplitude $A_{UT}^{\sin\psi}$ vanishes 
by definition. For $\pT  > 1.3$~GeV, the statistical uncertainties increase substantially 
with $\pT$. Here, there is an indication of an increase of the amplitude for pions, while 
for kaons it is compatible with zero within the uncertainties, apart from the point at 
the highest $\pT$, 
where the amplitude is 2.8 standard deviations above zero.  For negative hadrons 
the amplitudes are much smaller in magnitude, sometimes positive and sometimes negative, 
apart from the $\pi^-$ point at $\pT = 1.5$ GeV. Detailed investigations of  the data 
and the analysis leading to this exceptionally large asymmetry amplitude 
have not revealed any instrumental origin.

In the bottom panels of Fig.~\ref{fig:asym}, the measured $A_{UT}^{\sin\psi}$ amplitudes 
are presented as a function of $x_F$. For positive pions, the amplitudes are positive 
everywhere and increase nearly linearly with $x_F$ up to a value of approximately 0.06, 
with the exception of the point in the highest $x_F$ bin, where  the  
value is $0.10 \pm 0.01_{\text{stat}} \pm 0.01_{\text{sys}}$. 
For negative pions, the amplitude is  negative over most of the $x_F$ range and 
decreases linearly down to a value of about -0.04 for the last $x_F$ bin. 
These $x_F$ dependences of the pion asymmetry amplitudes look similar to the one 
observed in hadron-hadron collisions. For positive kaons, 
the amplitude is about constant around 0.07, with some small variation with $x_F$.
For negative kaons, the asymmetry amplitude is compatible with zero over most of the $x_F$ range, 
with a small positive excursion in the lowest $x_F$ bin, and a negative one
in the region around $x_F = 0.3$. 

\begin{figure}[t]
\centering
\plotje{.47\textwidth}{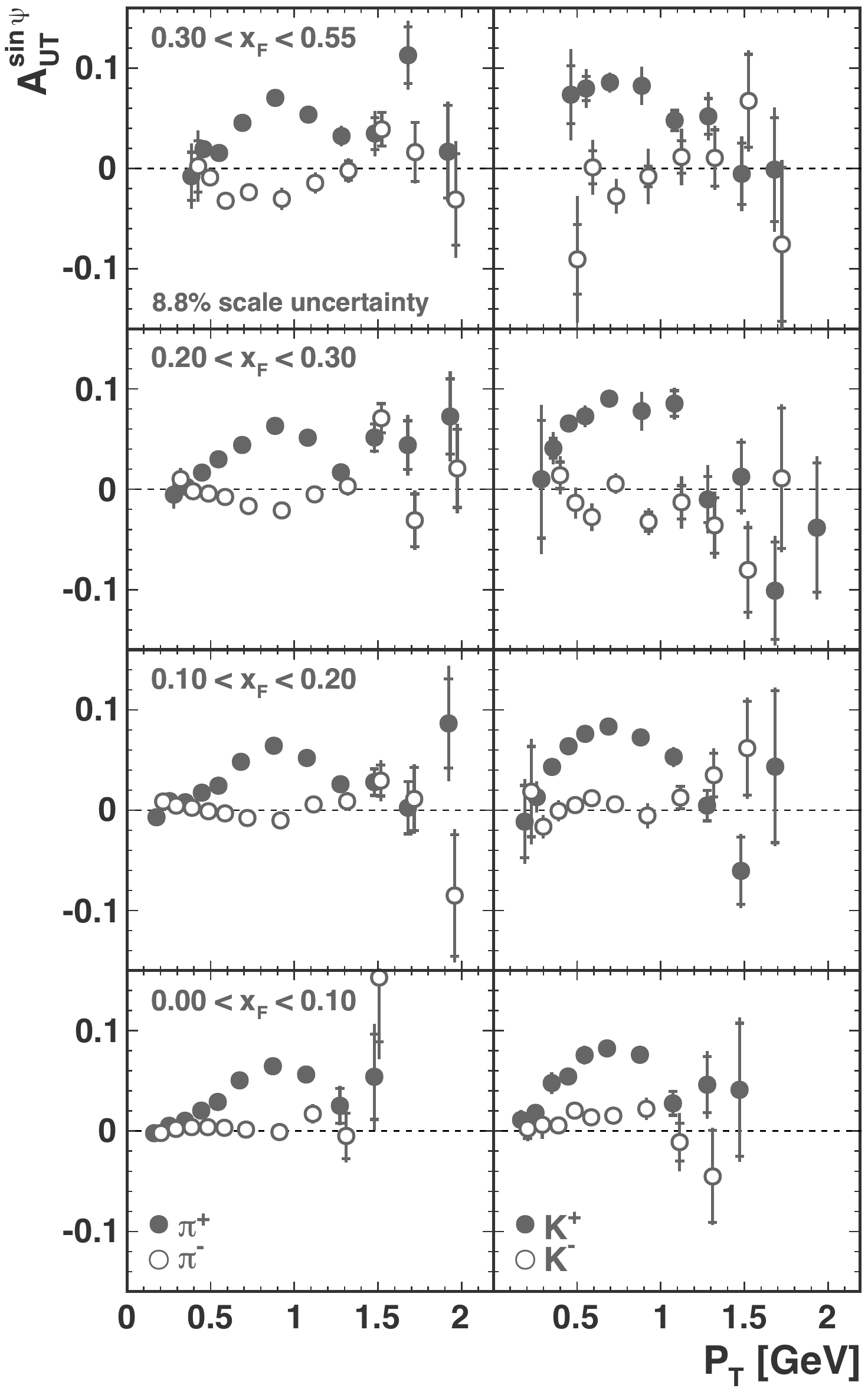} 
\caption{$A_{UT}^{\sin\psi}$ amplitudes for charged pions and kaons 
as a function of $\pT$ for various slices in $x_F$.
Symbol definitions and additional 8.8\% scale uncertainty as in Fig.~\ref{fig:asym}.
} 
\label{fig:asym2DpT}
\end{figure}

\begin{figure}[th]
\centering
\plotje{.47\textwidth}{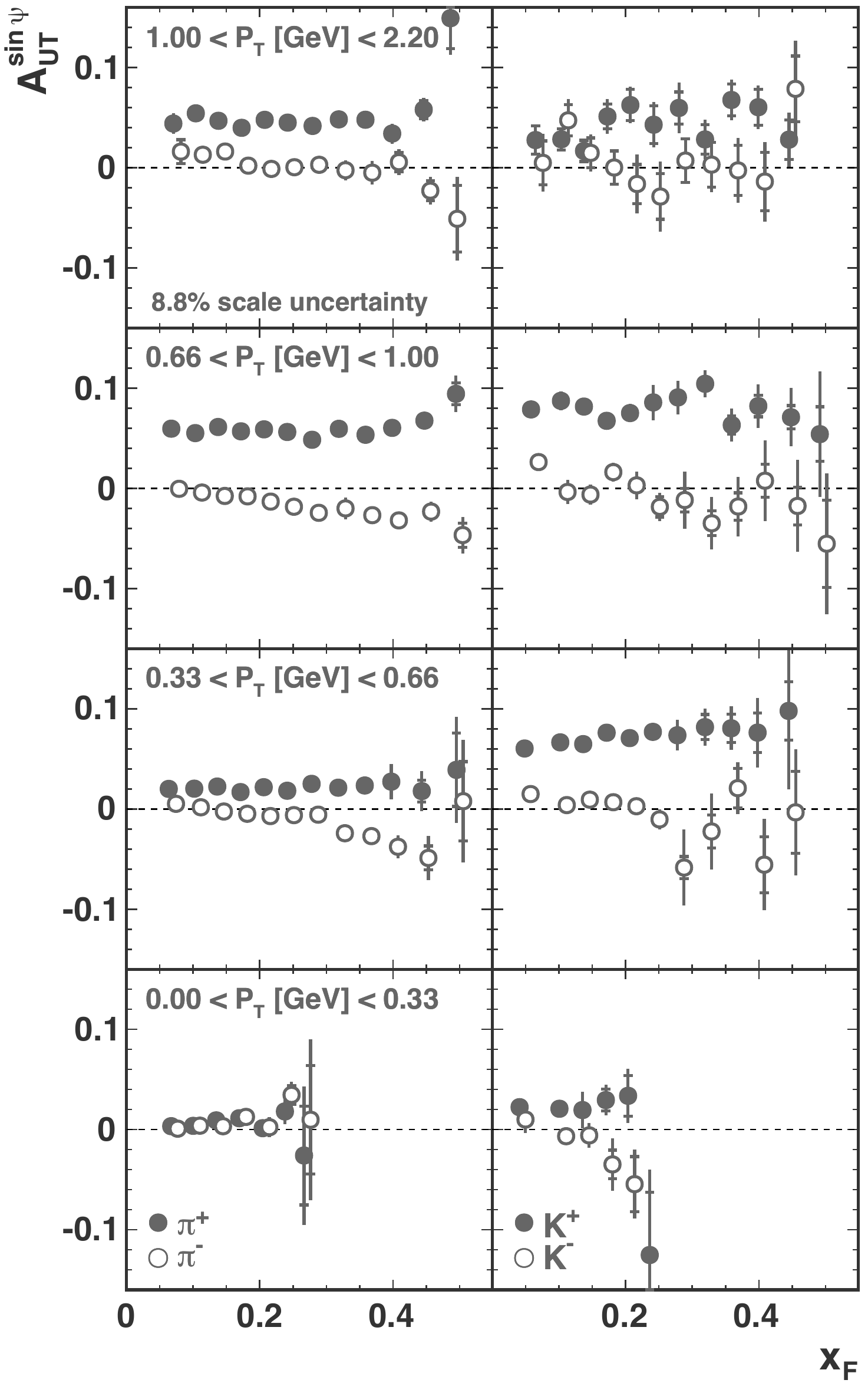} 
\caption{$A_{UT}^{\sin\psi}$ amplitudes for charged pions and kaons 
as a function of $x_F$ for various slices in $\pT$. 
Symbol definitions and additional 8.8\% scale uncertainty as in Fig.~\ref{fig:asym}.
} 
\label{fig:asym2DxF}
\end{figure}

The variables $x_F$ and $\pT$ are strongly correlated  in these measurements as can be 
seen from the bottom subpanels of Fig.~\ref{fig:asym}, where they are shown at the average 
bin kinematics. Hence, any observed kinematic dependence of $A_{UT}^{\sin\psi}$ cannot 
be uniquely ascribed to the variable plotted against but may stem from the 
underlying dependence on the kinematic variable over which the data are integrated. 
For this reason, a two-dimensional extraction of the asymmetry amplitudes was performed by 
binning simultaneously in $\pT$ and $x_F$. The resulting $A_{UT}^{\sin\psi}$ amplitudes 
are shown  as a function of  $\pT$ in four slices of $x_F$ in 
Fig.~\ref{fig:asym2DpT}, and in Fig.~\ref{fig:asym2DxF} as a function of $x_F$ in four 
slices of $\pT$. 
Only data points with a statistical uncertainty of the asymmetry 
amplitude smaller than 0.1 are shown. 
The $\pT$ dependence in the four $x_F$ slices is very similar in shape 
and magnitude, apart from increased statistical fluctuations.
For positive pions the amplitude is seen to be essentially independent of $x_F$ 
in all four slices in $\pT$. 
Therefore, it can be concluded that the apparent increase of the magnitude of 
the asymmetry amplitude with $x_F$ seen for positive pions in Figure~\ref{fig:asym} is just a 
reflection of the underlying dependence on $\pT$. 
In contrast, for negative pions the decrease with $x_F$ follows the one observed in the one-dimensional 
extraction. 
The dependence on $x_F$ of the kaon asymmetry amplitudes is less pronounced in the 
two-dimensional extraction, with a slight tendency towards an increase (decrease) 
with $x_F$ for positive (negative) kaons.
Note that in measurements of inclusive SSAs in proton-proton collisions, $A_N$ is 
seen to rise strongly for values of $x_F$ larger than about 0.3-0.4. 
For charged pions~\cite{brahms08} and neutral pions~\cite{star08} 
such an increase of $A_N$ with $x_F$ was seen even after 
binning the data in slices of $\pT$.

\begin{figure}[t]
\centering
\plotje{0.47\textwidth}{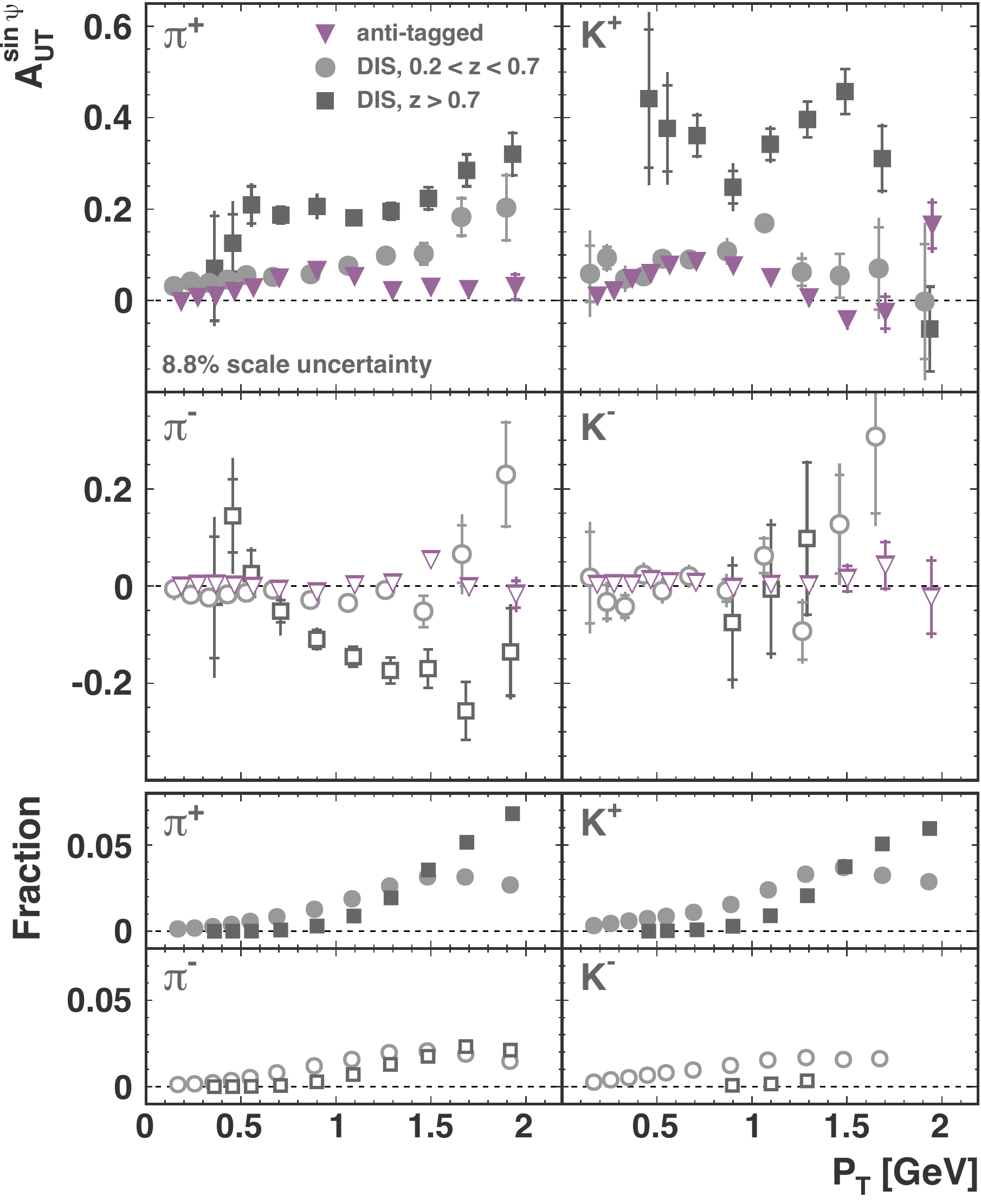} 
\caption{$A_{UT}^{\sin\psi}$ amplitudes for charged pions and kaons for the 
`anti-tagged' category  and the two DIS subsamples with $0.2 < z < 0.7$ and 
$z > 0.7$, respectively. Also shown are the relative fractions of the two DIS subsamples  
with respect to the total inclusive sample of the corresponding hadron species after 
correction for trigger efficiency. 
Positive (negative) particles are denoted by 
filled (open) symbols. Inner error bars 
show the statistical uncertainties and the total error bars represent statistical 
and systematic uncertainties added in quadrature. 
Not shown is an additional $8.8\%$  
scale uncertainty due to the precision of the measurement of the target polarization.
}
\label{fig:subsamples}
\end{figure}

In Figs.~\ref{fig:asym}--\ref{fig:subsamples}, the 
systematic uncertainties are added in quadrature to the statistical ones.
One contribution arises from different methods employed for the trigger-efficiency correction. 
An additional contribution, added in
quadrature to the previous one, arises from typical 
effects due to non-perfect experimental resolution and acceptance, and
is determined in a manner to include the effects of the necessary 
binning of finite statistics. This second contribution
was determined from a high-statistics Monte Carlo data sample obtained from a 
simulation using the 
program {\sc PYTHIA} 6.2 
~\cite{Sjostrand00, Sjostrand01}. This simulation~\cite{hermes10deltaG} contained a 
full description of the detector,  including effects such as acceptance,  
correction for particle deflection in 
the vertical target holding field, losses due to decay in flight and secondary strong interactions, and particle identification. In addition,
a spin-dependent azimuthal asymmetry was 
imposed on the simulated event sample according to Eq.~(\ref{eq:yields}). 
The functional form $\mathcal{A}_{UT,MC}^{\sin\psi}$ is a Taylor expansion in $\pT$ 
(up to fifth order) and $x_F$ (up to first order) around the average kinematics of the entire 
experimental data sample. 
The set of (up to) twelve parameters for each hadron species was obtained in a 
maximum-likelihood fit to the experimental data where the number of terms in the 
expansion was tuned to describe all measured asymmetry amplitudes. 
The $\sin\psi$ amplitudes $A_{UT,MC}^{\sin\psi}$ were then extracted 
from the now spin-dependent Monte Carlo sample
in the same way as described above for data.
The total systematic uncertainty in each bin corresponds to the maximum value of either 
the (in most bins negligibly small) statistical uncertainty in the Monte Carlo sample or 
the deviation between the model function $\mathcal{A}_{UT,MC}^{\sin\psi}$ evaluated at 
the average kinematics of the bin and the reconstructed $A_{UT,MC}^{\sin\psi}$ amplitude. 
This calculational approach is designed to address, among others, the differences between 
the asymmetry amplitudes evaluated at
the average values of their kinematical dependences and
the asymmetry amplitudes averaged over the bin range(s) of the
kinematical dependences. These differences can become large whenever
an amplitude's dependence deviates significantly from linear behavior over the
width of the bin in that dependence.
This makes the one-dimensional representation of Fig.~\ref{fig:asym}, 
where one integrates over the whole range in the variable not shown, 
more susceptible to systematic deviations. These can be 
observed, e.g., in the $x_{F}$ dependence of the $K^{+}$ 
asymmetry amplitudes, where the integration is over a
strongly varying $\pT$ dependence.
Additionally, the uncertainty on the measurement of the target polarization produces a 
$8.8\%$ scale uncertainty on the value of $A_{UT}^{\sin\psi}$ that is not included in the 
error bars/bands. Other possible sources of systematic uncertainty not included in the 
Monte Carlo simulation such as time-dependence of the measured 
amplitudes and the effect of different beam charges were found to be negligible. 


The inclusive data set 
presumably is a mixture of various contributions with different 
kinematic dependences. Therefore, it is difficult to draw conclusions about the underlying 
physics from the observed 
kinematic dependences of the inclusive asymmetry amplitudes. 
More insight may be gained by studying separately the asymmetries for 
the events without a scattered lepton in the acceptance (`anti-tagged' category) and 
the events with a scattered lepton in the acceptance (`tagged' or semi-inclusive category). These 
categories cover 
different kinematic regimes and are defined as follows: 

1) {\it `Anti-tagged' category}:
The undetected lepton in most cases 
had a small scattering angle and remained within the beam pipe. Hence $Q^2$ is small and 
$\pT$ is the only hard scale.  For these events, the difference between the transverse 
hadron momentum with respect to the beam direction, $\pT$, and with respect to 
the virtual-photon direction, $P_{h\perp}$, is small. The latter was used in the 
previous analyses of SSAs in semi-inclusive deep-inelastic scattering
~\cite{hermes05colsiv, hermes09siv,hermes10col}. The present data sample is dominated by 
the kinematic regime $Q^2\approx 0$~GeV$^2$ of quasireal photoproduction where the 
cross section is largest and where the hadronic component of the photon plays an 
important role. Generally speaking, in this kinematic range $l + p^\uparrow$ reactions 
are expected to be quite similar in nature to 
$h + p^\uparrow$ reactions. 
The `anti-tagged' category contains a small contamination of events at higher $Q^2$ 
where the electron is scattered into the horizontal gap of the spectrometer. 
These events amount to about one third in statistics of the 
semi-inclusive category, discussed below.  
Another tiny high-$Q^2$ contamination arises from lepton scattering angles
beyond the maximum polar angular acceptance of the spectrometer. These events 
occur dominantly at high $\pT$. 
Here, the large angle of the virtual photon with
respect to the beam axis often results in 
a significantly larger $\pT$ than $P_{h\perp}$ of the hadrons.
After correction for trigger efficiency, about $98\%$ of all hadrons belong to 
the `anti-tagged' category. The fraction of these hadrons with respect to the total 
inclusive sample is nearly $100\%$ at low $\pT$. It decreases 
monotonically to about 
$85-90\%$ for positive hadrons
and to more than $90\%$ for negative hadrons at the highest $\pT$ values.

\begin{table}
\begin{center}
\begin{tabular}{|c||c|c|c|c|c|}
\hline
subsample &  $\pi^+$ & $\pi^-$    &  K$^+$   & K$^-$  \\
\hline
`anti-tagged'  &  170.5 & 140.7 & 14.3  & 7.2  \\
\hline
`tagged'  &  1.93 & 1.49 & 0.26  & 0.13  \\
\hline
DIS, $0.2 < z <0.7$ &  0.69 & 0.49 & 0.12  & 0.05  \\
\hline
DIS, $z > 0.7$ &  0.061 & 0.037 & 0.013  & 0.001  \\
\hline
\end{tabular}
\caption{Accumulated yields of hadrons (in million) for the various event samples without 
and with an electron in the spectrometer acceptance, after correction for trigger efficiency. 
The DIS subsamples are part of the `tagged' category, as explained in the text.
}
\label{counts}
\end{center}
\end{table}

2) {\it `Tagged' or semi-inclusive category:} 
The scattered positron was recorded in the spectrometer acceptance
and kinematic quantities like $y$, $z$, $Q^2$, $x$, and $W^2$ could be determined, where in 
the laboratory system $y \equiv P \cdot q/(P \cdot k)$ is the fractional beam energy carried 
by the virtual photon and $z \equiv P \cdot P_h/(P\cdot  q)$ is the fractional 
virtual-photon energy carried by the hadron. The quantity  $x \equiv Q^2/(2P \cdot q)$  is 
the Bjorken scaling variable with $-Q^2 = q^2 \equiv (k-k')^2$, and $W^2 \equiv (P + q)^2$ 
is the squared invariant mass of the virtual-photon nucleon system. Here, $P$, $k$, $k'$, 
and $P_h$ are the four-momenta of the target nucleon, the incident and scattered lepton, and 
the produced hadron, respectively.
This category can be further divided into several subsamples covering different 
kinematic regions.

For the present analysis, two of these subsamples have been selected 
that affect substantially the observed asymmetries at large $\pT$: 

\noindent - {\it DIS events with $0.2 < z < 0.7$:}
This subsample is identical to the one used previously
~\cite{hermes05colsiv, hermes09siv,hermes10col} for the determination
of azimuthal transverse single-spin asymmetries in semi-inclusive deep-inelastic 
scattering related to the Sivers and transversity distributions and the 
Collins fragmentation function. Here, events were selected according to the 
kinematic requirements $Q^2 > 1$~GeV$^2$, $W^2 > 10$~GeV$^2$, $0.023 < x < 0.4$, and 
$0.1 < y < 0.95$. The fractional hadron energy was required to be in the range  
$0.2 < z < 0.7$. For this sample, $\langle Q^2 \rangle$ rises from $\sim 2.2$~GeV$^2$ 
at low $\pT$ to $\sim 4.3$~GeV$^2$ at high $\pT$, and $\langle Q^2 \rangle$ is always 
larger than {\bf $ \langle \pT^2 \rangle$} apart from the two highest $\pT$ bins; 

\noindent - {\it DIS events with $z > 0.7$:}
The kinematic requirements are identical to those of the above subsample, apart from 
the requirement for the fractional hadron energy. Only hadrons with $z > 0.7$ are selected. 
The average value of $Q^2$ rises from
$\sim 1.5$~GeV$^2$ to $\sim5.5$~GeV$^2$ and 
$\langle Q^2 \rangle > \langle  \pT^2 \rangle$ over the whole  $\pT$ range.

The total number of hadron tracks in the `anti-tagged' and the `tagged' categories and in 
the two DIS subsamples is listed in Table~\ref{counts}. 
The remaining events of the `tagged' category contribute only at $x_F < 0.2$ and 
$\pT < 0.9$~GeV and will not be discussed further.

In Figure~\ref{fig:subsamples}, the $A_{UT}^{\sin\psi}$ amplitudes  are presented as a 
function of $\pT$ for the `anti-tagged' category and 
the two DIS subsamples with $0.2 < z < 0.7$ and $z > 0.7$, respectively. Also shown are 
the relative fractions of these two subsamples with respect to the total inclusive sample 
of the corresponding hadron species after correction for trigger efficiency.
The relative fractions are generally larger for positive hadrons than for negative hadrons. 
For $\pT < 1$~GeV, the fractions are below 1\%. 
In the highest two $\pT$ bins, the fraction of DIS events 
with $z > 0.7$ dominates for positive hadrons and reaches values of about 6\%, while the 
DIS contribution with $0.2 < z < 0.7$ stays below 4\%. 
As can be seen from Figure~\ref{fig:subsamples}, the asymmetry amplitudes for the `anti-tagged' category
and for the two subsamples of the `tagged' category show several remarkable peculiarities:

\noindent -  {\it `Anti-tagged' category}:  
The asymmetry amplitudes are, over most of the $\pT$ 
range, essentially identical to the inclusive 
amplitudes as expected from the fact that 
this sample comprises about $98\%$ of the whole statistics. One can therefore safely 
conclude that essentially all of the kinematic dependences of the inclusive data set 
observed for $\pT$ below approximately 1.5 GeV originate from quasi-real photoproduction. 
Since $\pT$ is the only hard scale, the origin of the asymmetries can most likely be 
explained by higher-twist contributions. At low values of $\pT$, where one observes a 
rise of the asymmetry amplitudes for positive hadrons, $\pT$ is comparable to $\Lambda_\text{QCD}$ and 
theory cannot presently make reliable predictions about the magnitude and $\pT$ dependence 
of the amplitudes. At high $\pT$, the `anti-tagged' asymmetry amplitude is consistently 
smaller than the inclusive amplitude
for positive pions and its $\pT$ dependence is, 
within uncertainties, compatible with a constant or a decrease with $\pT$ as one would 
expect for this class of events~\cite{anselmino10,Kang11b}. 
At $\pT > 1.3$~GeV the contributions from the other subsamples become sizable causing 
the increase with $\pT$ observed for the inclusive asymmetry amplitude.

\noindent - {\it DIS events with $0.2 < z < 0.7$}: For positive pions, $A_{UT}^{\sin\psi}$ 
is positive and larger than the `anti-tagged' amplitude. It rises rather linearly with 
$\pT$ from a value of approximately 0.04 at low $\pT$  to approximately 0.2 at the highest 
$\pT$ values, where the statistical uncertainties are rather large. For negative pions, 
the amplitude is (apart from the two highest $\pT$ points) consistently negative and larger 
in magnitude than the asymmetry amplitude for the `anti-tagged' sample over the whole range of $\pT$. 
As stated above, $Q^2$ is the largest scale over essentially the whole $\pT$ range 
and transverse-momentum-dependent distribution and fragmentation functions can contribute 
without $\pT$-suppression. Since the angle $\psi$\ and the Sivers angle $\phi - \phi_s$ 
are closely related, one can expect that the observed $\pT$ dependence is predominantly caused 
by the Sivers effect. In fact, the asymmetries are very similar to those in 
Ref.~\cite{hermes09siv}, where it was concluded that the small amplitudes for $\pi^-$ 
require cancellation effects, e.g., from a down-quark Sivers function opposite in sign to 
the dominant up-quark Sivers function.

\noindent - {\it DIS events with $z > 0.7$}:
Large asymmetries are observed for this subsample for both pion charges 
and especially for positive kaons, where 
the amplitudes reach values of more than 0.4. 
For positive pions the 
amplitude is rather constant with a value of around 0.2 in the 
$\pT$ range 0.5--1.5 GeV, and rises up to a value above 0.3 at the highest $\pT$ bin. 
For negative pions, the amplitude is negative and decreases from 
approximately zero at $\pT \sim 0.5$~GeV down to a  value of about -0.2 at high $\pT$. 
This subsample receives contributions from processes that 
can become significant only in this kinematic region. Pions 
receive contributions from decays of exclusive mesons like, e.g., the
$\rho$ meson~\cite{hermes09rho} that can contribute up to about $50\%$ ($30\%$) to the
yield of $\pi^-$ ($\pi^+$) at large $z$~\cite{hermes12mul}. For kaons, the corresponding contributions from 
$\phi$ decays are less than 10\%. For positive pions there is in addition a contribution 
from exclusive production, $l p \rightarrow l' \pi^+ n$, which  has, however, been 
measured~\cite{hermes10pi+} to constitute only approximately $3\%$ of this sample. 
The corresponding contributions for the quasi-exclusive production of negative pions from  
$l p \rightarrow l' \pi^- \Delta^{++}$ or positive kaons from 
$l p \rightarrow l' K^+ \Lambda$ are expected to be even smaller~\cite{Diehl05klambda} and 
no such quasi-exclusive channel exists for negative kaons. The large asymmetry 
amplitude seen 
for negative pions may indicate that a large fraction of events in this subsample stems 
from the favoured fragmentation of the struck quark (here the down quark) and that 
the asymmetry possibly preserves information from the down-quark Sivers
function without dilution from disfavoured fragmentation of the otherwise dominating up quark. 
Indeed, the signs and relative magnitudes of the pion and kaon asymmetry amplitudes observed
are not inconsistent with the values of the up and down quark Sivers functions extracted in
phenomenological fits ~\cite{Anselmino11}.

In summary, transverse azimuthal single-spin asymmetries are measured in 
inclusive and semi-inclusive electroproduction of charged pions and kaons.  A 
two-dimensional extraction of the asymmetry amplitudes is performed by binning 
simultaneously in the component of the hadron-momentum transverse to the incoming lepton beam, 
$\pT$, and the Feynman-$x$ variable, $x_F$. For positive pions, the resulting amplitudes 
are found to be essentially independent of $x_F$. The 
apparent increase with $x_F$ after integration over $\pT$ is mostly a reflection of 
the underlying dependence on $\pT$. For negative pions, and less significantly for negative 
(positive) kaons, the asymmetry amplitudes decrease (increase) with $x_F$, 
also in the case of a two-dimensional extraction.
The amplitudes as a function of $\pT$ are positive for 
the positive hadrons and slightly larger for $K^+$ compared to $\pi^+$. 
They rise smoothly 
with $\pT$ from zero at low $\pT$ up to a maximum value of approximately 0.06 (0.08) for 
pions (kaons) at  $\pT \simeq 0.8$~GeV and then decrease with increasing $\pT$. 
The data sample is dominated by the kinematic regime $Q^2\approx 0$~GeV$^2$ of 
quasi-real photoproduction, where $\pT$ is the only hard scale.  The origin of the 
observed asymmetries can, therefore, most likely be explained by higher-twist contributions. 
At $\pT$ values above 1.5~GeV there are sizable contributions from events with an electron 
in the acceptance and large values of $Q^2$. The asymmetries for the subsample within 
DIS kinematics and fractional energies of the hadron in the range $0.2 < z < 0.7$ 
can likely be related to the transverse-momentum dependent Sivers distribution function. 
Very large asymmetry amplitudes are observed for positive pions and kaons 
and negative pions for the DIS subsample with high values of the fractional hadron energy $z$. In this 
kinematic regime exclusive processes can contribute substantially to the asymmetry and 
effects from the favoured fragmentation
of the struck quark dominate. The data may be very helpful in formulating a 
better understanding of spin-orbit effects of partons within the nucleon.

We gratefully acknowledge the {\sc DESY} management for its support and the staff
at {\sc DESY} and the collaborating institutions for their significant effort.
This work was supported by 
the Ministry of Economy and the Ministry of Education and Science of Armenia;
the FWO-Flanders and IWT, Belgium;
the Natural Sciences and Engineering Research Council of Canada;
the National Natural Science Foundation of China;
the Alexander von Humboldt Stiftung,
the German Bundesministerium f\"ur Bildung und Forschung (BMBF), and
the Deutsche Forschungsgemeinschaft (DFG);
the Italian Istituto Nazionale di Fisica Nucleare (INFN);
the MEXT, JSPS, and G-COE of Japan;
the Dutch Foundation for Fundamenteel Onderzoek der Materie (FOM);
the Russian Academy of Science and the Russian Federal Agency for 
Science and Innovations;
the Basque Foundation for Science (IKERBASQUE) and the UPV/EHU under program UFI 11/55;
the U.K.~Engineering and Physical Sciences Research Council, 
the Science and Technology Facilities Council,
and the Scottish Universities Physics Alliance;
the U.S.~Department of Energy (DOE) and the National Science Foundation (NSF);
as well as the European Community Research Infrastructure Integrating Activity
under the FP7 "Study of strongly interacting matter (HadronPhysics3, Grant
Agreement number 283286)".






\bibliographystyle{model1a-num-names_noDOI}
\bibliography{dc89}







\end{document}